\documentclass{aa} 

\usepackage{graphicx,natbib} 
\bibpunct{(}{)}{;}{a}{}{,} 

\newcommand{\solarlum}{${\rm L}_{\sun}$~} 
\newcommand{\solarmass}{${\rm M}_{\sun}$~} 
\newcommand{\solarmassyr}{\solarmass ${\rm yr}^{-1}$~} 
\newcommand{\Eup}{${\rm E_{up}}$~} 
\newcommand{\Mdot}{$\dot{\rm M}$~} 
\newcommand{\Mstar}{${\rm M_{*}}$~} 
\newcommand{\Xin}{$X_{in}$~} 
\newcommand{\Xout}{$X_{out}$~} 
 
\begin{document} 
 
\date{\today} 
 
\title{Water emission in NGC1333-IRAS4\thanks{Based on observations 
     with ISO, an ESA project with instruments funded by ESA Member 
     States (especially the PI countries: France, Germany, the 
     Netherlands and the United Kingdom) with the participation of ISAS 
     and NASA.}} 
 
\subtitle{The physical structure of the envelope} 
 
\titlerunning{Water emission in NGC1333-IRAS4} 
 
\author{S. Maret\inst{1} \and C. Ceccarelli\inst{2,3} \and 
   E. Caux\inst{1} \and A.G.G.M. Tielens\inst{4} \and 
   A. Castets\inst{2}} 
 
\institute{Centre d'Etude Spatiale des Rayonnements, CESR/CNRS-UPS, BP 
    4346, F-31028 Toulouse Cedex 04, France \and Observatoire de 
    Bordeaux, BP 89, F-33270 Floirac, France \and Laboratoire 
    d'Astrophysique, Observatoire de Grenoble, B.P. 53, F-38041 
    Grenoble Cedex 09, France \and Space Research Organization of the 
    Netherlands, P.O. Box 800, 9700 AV Groningen, The Netherlands} 
 
\offprints{S\'ebastien Maret, \email{sebastien.maret@cesr.fr}} 
 
\date{Received {\today} /Accepted} 
 
\abstract{We report ISO-LWS far infrared observations of CO, water and
     oxygen lines towards the protobinary system IRAS4 in the NGC1333
     cloud. We detected several water, OH, CO rotational lines, and
     two [OI] and [CII] fine structure lines. Given the relatively
     poor spectral and spatial resolution of these observations,
     assessing the origin of the observed emission is not
     straightforward.  In this paper, we focus on the water line
     emission and explore the hypothesis that it originates in the
     envelopes that surround the two protostars, IRAS4 A and B, thanks
     to an accurate model. The model reproduces quite well the
     observed water line fluxes, predicting a density profile, mass
     accretion rate, central mass, and water abundance profile in
     agreement with previous works.  We hence conclude that the
     emission from the envelopes is a viable explanation for the
     observed water emission, although we cannot totally rule out the
     alternative that the observed water emission originates in the
     outflow.  The envelopes are formed by a static envelope where the
     density follows the $r^{-2}$ law, at $r \geq 1500$ AU, and a
     collapsing envelope where the density follows the $r^{-3/2}$ law.
     The density of the envelopes at 1500 AU from the center is $\sim
     4 \times 10^6$ cm$^{-3}$ and the dust temperature is $\sim 30$ K,
     i.e. about the evaporation temperature of CO-rich ices.  This may
     explain previous observations that claimed a factor of 10
     depletion of CO in IRAS4, as those observations probe the outer
     $\leq 30$ K region of the envelope. The water is $\sim 5 \times
     10^{-7}$ less abundant than H$_2$ in the outer and cold envelope,
     whereas its abundance jumps to $\sim 5 \times 10^{-6}$ in the
     innermost warm region, at $r\leq 80$ AU where the dust
     temperature exceeds 100 K, the evaporation temperature of
     H$_2$O-rich ices.  We derive a mass of 0.5 \solarmass for each
     protostar, and an accretion rate of $5 \times 10^{-5}$
     \solarmassyr, implying an age of about 10000 years, if the
     accretion rate remains constant.  We finally discuss the
     difference between IRAS4 and IRAS16293-2422, where a similar
     analysis has been carried out. We found that IRAS4 is probably a
     younger system than IRAS16293-2422.  This fact, coupled with the
     larger distance of IRAS4 from the Sun, fully explains the apparent
     difference in the molecular emission of these two sources, which
     is much richer in IRAS16293-2422.
}
 
\maketitle 
 
 
\section{Introduction} 
 
The south part of the NGC1333 reflection nebulae, in the Perseus
cloud, is an active star forming region, containing many infrared
sources associated with molecular flows and numerous Herbig-Haro
objects.  IRAS4 was first identified by \citet{Jennings86}, and
further observations \citep{Sandell91} revealed IRAS4 it was a binary
system resolved into two components, named IRAS4A and IRAS4B, and
separated by 31$\arcsec$. Interferometric observations
\citep{Lay95,Looney00} have shown further multiplicity of the two
sources.  IRAS4A is itself a binary system with a separation of
10\arcsec, and there is some evidences that IRAS4B could also be a
multiple system, with a separation of 0.5\arcsec.
 
The distance of the NGC1333 cloud is much debated. \citet{Herbig83}
found a distance of 350 pc for the Perseus OB2 association \citep[a
more recent estimate based on the Hipparcos data gives
318$\pm$27;][]{deZeeuw99}, but extinction observations towards NGC1333
itself \citep{Cernis90} suggest that it may be as close as 220 pc.
Assuming a distance of 350 pc, \citet{Sandell91} measured a system
total luminosity of 28 \solarlum (11 \solarlum at 220 pc) equally
shared between IRAS4A and B.  They derived an envelope mass of 9 and 4
\solarmass respectively (3.5 and 1.5 \solarmass at 220 pc). This
relatively large mass, together with the low bolometric luminosity
suggest that both sources are deeply embedded and probably very young.
They have been classified as \emph{Class 0} sources \citep{Andre93}.
IRAS4A and B are both associated with molecular outflows, detected in
CO, CS \citep{Blake95} and SiO \citep{Lefloch98} millimeter
transitions.  The outflow originating from IRAS4A is very highly
collimated, whereas that originating from IRAS4B is rather compact and
unresolved in single dish observations \citep{Knee00}.  The dynamical
ages of both outflows are a few thousands years.
 
In the past years, many observational studies have been focused on the
continuum emission of IRAS4.  Recent works include maps of the region
obtained with IRAM at 1.3 mm \citep{Lefloch98} and with SCUBA at 450
and 850 $\mu$m \citep{Sandell01}.  An accurate modeling of the
continuum emission has been very recently carried out by \citet[][
hereafter JSD02]{Jorgensen02}, who reconstructed the dust temperature
and density profiles across the two envelopes.
 
The molecular line emission is probably a better and certainly a
complementary tool to probe the dynamical, chemical and physical
structure of the envelopes of IRAS4.  The last decade has seen
flourishing several studies of molecular line profiles
\citep[e.g.][]{Gregersen97, Evans99} and line spectra \citep{Blake95},
all having in common the goal of reconstructing the physical structure
of the protostellar envelopes.  Specifically, \citet{Blake95} carried
out a multifrequency study of several molecules in IRAS4, including
H$_2$CO and CH$_3$OH. Their two major results regarding the structure
of the IRAS4 envelopes are: 1) a large depletion, around a factor
10-20, of CO and all molecules in the envelope, and 2) the presence of
a region with an increased abundance of CS, SiO and CH$_{3}$OH, that
the authors attribute to mantles desorption caused by grain-grain
collisions induced by the outflows originating from the two
protostars.  More recently interferometric observations by
\citet{DiFrancesco01} \citep[see also][]{Choi99} detected an inverse
P-cygni profile of the H$_{2}$CO $3_{2,1}-2_{1,1}$ line on a $2''$
scale towards both IRAS4A and B, providing the least ambiguous
evidence of infall motion towards a protostar ever.  From a simple
two-layer modeling, they derived an accretion rate of $1.1 \times
10^{-4}$ and $3.7 \times 10^{-5}$ \solarmassyr, an inner mass of 0.7
and 0.2 \solarmass, and an age of 6500 and 6200 yr (assuming constant
accretion rate) for IRAS4A and IRAS4B respectively.
 
In this paper we concentrate on the far infrared (FIR) line spectrum,
and in particular the water line spectrum observed with the {\it Long
Wavelength Spectrometer} \citep[][ herein after LWS]{Clegg96} on board
ISO \citep{Kessler96} in the direction of IRAS4.  The goal of this
study is to check whether the observed water line emission can be
attributed to the thermal emission of the envelopes surrounding the
IRAS4 protobinary system.  Water lines have in fact been predicted to
be a major coolant of the gas in the collapsing envelopes of low-mass
protostars (Ceccarelli et al. 1996, hereafter CHT96; Doty \& Neufeld
1997).  Given the relatively large range of level energies (from $\sim
100$ to $\sim 500$ K) and spontaneous emission coefficients (from
$10^{-2}$ to $\sim 1$ s$^{-1}$) of the water transitions observed by
ISO-LWS, the observed lines can in particular probe the innermost
regions of the envelope.  This makes the analysis of the ISO-LWS water
lines a precious and almost unique tool (when considering the water
abundance across the envelope).  The reverse of the coin is that
assessing the actual origin of the water emission is somewhat
difficult and still debated, as the spectral and spatial resolutions
of ISO-LWS are relatively poor to disentangle the various components
falling into the beam.  For example, strong molecular line emission is
often associated with the outflows emanating from young protostars
\citep[e.g.][]{Bachiller97}.  As already mentioned, the line emission
from CO, CS and other molecules are certainly contaminated by the
outflowing gas in IRAS4.  Nonetheless, low lying lines seem to be more
affected than high lying lines in first instance, and different
molecules suffer differently from this ``contamination'', as proved by
the \citet{DiFrancesco01} observations.  Although water has been
predicted to be very abundant in shocked gas, the published ISO
observations show that the water emission is usually stronger towards
the central sources and weaker, if detected at all, in the direction
of the peaks of the outflows powered by low mass protostars
\citep[see][ for a review]{Ceccarelli00a}. When water lines are
detected in clear-cut shocked regions, the water abundance seems to be
lower than that predicted by the models, like in the case of HH54
\citep{Liseau96} or HH7-11 \citep{Molinari99, Molinari00}, or in the
outflows of IRAS4 (see next section). Finally, SWAS observations seem
to support the evidence that the water abundance in the shocked
regions is a few times $10^{-6}$ \citep{Neufeld00}. These facts,
together with the apparent correlation between the observed water
emission and the 1.3 mm continuum, and the lack of correlation with
SiO emission\footnote{SiO emission is usually associated with the
outflow strong shocks \citep[e.g.][]{Bachiller97}, and it is believed
to be a product of grain mantle desorption
\citep{Caselli97,Schilke97}} in low mass protostars
\citep{Ceccarelli99} play in favor of a relatively low contamination
of the ISO-LWS observed water emission by the outflow and encourage us
to explore in detail this hypothesis for the IRAS4.
 
In the specific case of IRAS4, the {\it Submillimiter Wavelengths
Astronomical Satellite} \citep[SWAS;][]{Melnick00} observed the ground
o-H$_{2}$O line at 557 GHz (Neufeld et al. 2000, Bergin et al. 2002).
Given its relatively large linewidth ($\sim$18 km s$^{-1}$) the 557
GHz line is certainly dominated by the outflow emission.  Nonetheless,
this does not imply that the ISO FIR water lines also originate in the
outflow, and this for two reasons.  First, the beamwidth ($\sim 4'$)
of the SWAS observations, being about 3 times that of ISO-LWS,
encompasses the entire outflow, whereas the ISO observations do not
encompass the two emission peaks of the outflow (see also
\ref{origin}), but only the envelope.  Second, the 557 GHz transition,
being the water ground transition, is more easily excited than the FIR
water lines, and therefore the latter probably probe different
regions. In fact, Bergin et al. (2002) find that most of the 557 GHz
line must originate in a component colder, hence different, by that
probed by the FIR water lines, even under the assumption that they
probe the outflow.  To summarize, decicing whether the observed FIR
water emission in IRAS4 originates in the outflow or in the envelope
remains an open question, based on the available present observations. In
this article we explore in detail the latter hypothesis and submit it
to the scrutiny of an accurate modeling, trying hence to answer to the
question on a theoretical basis.  At this scope we used the CHT96
model, already successfully applied to the solar type protostar
IRAS16293-2422 which allowed to explain more than two dozen observed
ISO-LWS water lines and ground-based millimeter SiO and H$_2$CO lines
\citep{Ceccarelli00a,Ceccarelli00b}. One of the major results of that
work is the prediction of the existence of a hot core like region in
the innermost part of the envelope of IRAS16293-2422, in which the
dust temperature exceeds the evaporation temperature of interstellar
ice ($\simeq$ 100 K). These studies have been confirmed by the recent
analysis by \citet{Schoier02} of several other molecular transitions.
Such hot cores are well studied around massive protostars where --
driven by reactions among the evaporated ice molecules in the warm gas
-- their chemical composition differs substantially from that of
quiescent clouds \citep{Walmsley89, Charnley92}.  Hot cores around low
mass protostars may actually have a different chemical composition
\citep{Ceccarelli00b}. This molecular complexity may be of
prime interest on account of a possible link to the chemical history
of the solar nebula and hence the molecular inventory available to the
forming Earth and other solar system planets and satellites.
 
In order to understand the physical and chemical processes that take
place during the first stages of star formation, it would be necessary
to undertake a work similar to that the one done on IRAS16293-2422 on
a larger sample of protostars.  In this paper we present a study of
the structure of the envelope of NGC1333-IRAS4, obtained using ISO-LWS
observations of the H$_2$O far-infrared lines.  A preliminary analysis
of the same set of data has already been presented in
\citet{Ceccarelli99} and \citet{Caux99b}.  Here we revisit the data
using a new calibration and compare the observations with the CHT96
model predictions, testing a large range of model parameters.  This
study is part of a large project aimed to model the water emission in
several low mass protostars. The water observations are complemented
with formaldehyde and methanol ground based observations, to have a
complete budget of the most abundant molecules in the innermost
regions of the protostellar envelopes (Maret et al. in preparation).
Finally, the structure obtained by the analysis of these observations
will be compared with that independently obtained by continuum
observations by JSD02.
 
The outline of the article is the following.  In \S \ref{observations}
we present the data, in \S \ref{modeling} we describe the modeling of
the observed lines and in \S \ref{discussion} we discuss the physical
and chemical structure of the envelope, namely the density and
temperature profiles, as well as the abundances of the major species
across the envelope.  Besides, the central mass of the protostar and
its accretion rate can also be constrained by these observations and
modeling, yielding an alternative method to measure these two key
parameters.  In \S \ref{discussion} we compare the results of the
present study with previous studies of IRAS4.  Finally, we discuss the
similarities and differences between IRAS4 and IRAS16293-2422, and
highlight the importance of complementary ground-based, higher spatial
and spectral resolution observations to understand the physical and
chemical processes taking place in the innermost regions of low-mass
envelopes.
 
\section{Observations and results} 
\label{observations} 
 
A full-range spectral survey (43-196 $\mu$m) of the IRAS4 region was
performed using LWS. The observations were obtained on three positions
in LWS grating mode, with a spectral resolution of about 200 (AOT
LO1). The first position was centered in between IRAS4A and IRAS4B
($\alpha_{2000}\,=\,03^{\rm h}29^{\rm m}11.9^{\rm s}$,
$\delta_{2000}\,=\,31 \degr 13 \arcmin 20.3 \arcsec$), and hence LWS
80$\arcsec$ beam includes both sources. The two other positions aimed
to the lobe peaks of the outflow powered by IRAS4: NE-red,
$\alpha_{2000}\,=\,03^{\rm h}29^{\rm m}15.6^{\rm s}$,
$\delta_{2000}\,=\,31 \degr 14 \arcmin 40.1 \arcsec$), and SW-blue,
$\alpha_{2000}\,=\,03^{\rm h}29^{\rm m}06.6^{\rm s}$,
$\delta_{2000}\,=\,31 \degr 12\arcmin 08.7 \arcsec$). These
observations, performed during revolution 847, are made of 30 scans on
the central position and 10 scans on the two other positions. The
sampling rate was 1/4 of the grating resolution element (0.29 $\mu$m
in the 43-92 $\mu$m range, and 0.15 $\mu$m in the 84-196 $\mu$m
range). The integration time for each sampled point was 12 sec on the
central position, and 4 sec on the outflow positions.
 
\begin{figure*} 
    \centering \includegraphics[width=17cm]{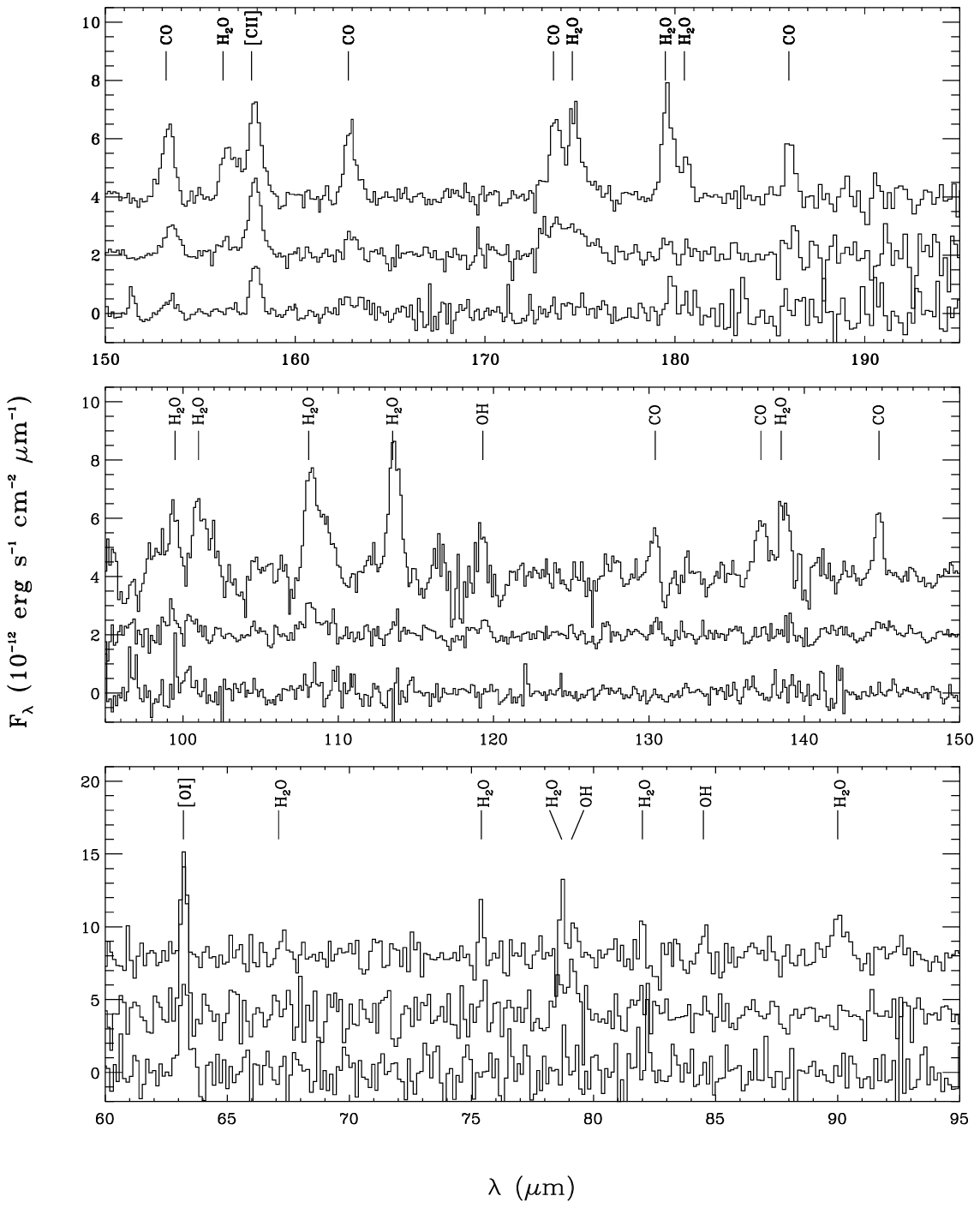}
    \caption{ISO-LWS spectra observed towards IRAS4 on source (top
      line), NE-red (middle line) and SW-blue (bottom line). }
    \label{spectrum} 
\end{figure*} 
 
The same data set has been previously analyzed by \citet{Ceccarelli99}
and \citet[][ hereafter GNL01]{Giannini01}. We here report again the
analysis of this dataset as we used an improved data processing
module, allowing the correction of the transient effects that affect
the LWS detectors \citep{Caux01}. As for standard pipeline products,
the data are calibrated against Uranus, and the calibration
uncertainty is estimated to be better than 30\% \citep{Swinyard98}.
On the three observed positions, the spectra were then defringed for
all ten detectors, and the continuum was removed fitting a polynomial
baseline outside the lines. The data were then averaged over the ten
detectors and binned at a single resolution to produce a single
spectrum for each observed position.  The line flux measurements were
finally performed with the ISAP package, using gaussian fits. A
particular attention was given to the determination of uncertainties
associated with the line fluxes measurement. These are due to the
statistical uncertainties, the absolute calibration and the baseline
determination uncertainties that affect low resolution LWS
observations. In grating mode, on a spectrum rich in lines as it is
the case here, the line confusion may be important, making the
baseline determination difficult. This uncertainty, often neglected in
the literature, should be taken into account, as it can lead to
important errors, especially for faint lines.
 
Figure \ref{spectrum} shows the observed 60-200 $\mu$m spectra in the
three observed positions, and Table 1 reports the measured line
fluxes. The errors quoted in this table include statistical errors,
errors due to the uncertainty of the baseline removal, and an absolute
calibration error of 30\%.  The first striking result of these
observations is the dramatic difference within the three spectra:
while that including IRAS4A and 4B is very rich in CO and water lines,
molecular emission is barely detected towards the outflow peaks (where
the millimeter CO emission is the brightest). On the contrary, the
fine structure [OI] 63$\mu$m and [CII] 157$\mu$m lines have comparable
fluxes in the three observed positions.  Finally, we wish to comment
our results with respect to previous published data reductions.  The
present line flux determination agrees with that quoted by
\citet{Ceccarelli99} when considering the uncertainties. However, we
note several differences with respect to the values quoted in
GNL01. While there is a relatively good agreement between their fluxes
of the strongest lines and ours (see Table \ref{fluxes}), there is a
noticeable discrepancy between our respective reductions regarding the
weakest lines. We think that this is probably due to a too optimistic
evaluation of the noise in GNL01.  For example, in the NE-red position
we find a statistical error around 175 $\mu$m of 1.5$\times$
10$^{-13}$ erg s$^{-1}$ cm$^{-2}$ $\mu$m$^{-1}$, while Giannini et al.
quote 2 $\times$ 10$^{-14}$ erg s$^{-1}$ cm$^{-2}$ $\mu$m$^{-1}$ in
their Table 3.  We do not confirm neither the detection of CO lines
with J$_{up} \geq 21$\footnote{CO 21-20 is definitively undetected, CO
22-21 is blended with OH at 119 $\mu$m, while CO 23-22 and CO 24-23
are blended with water lines at 113.5 and 108.8 $\mu$m line
respectively. One can note that the 108 $\mu$m is wider than expected,
which could be due to the presence of the CO 24-23 line. Nonetheless,
this detection would be inconsistent with the non detection of lower
and higher J CO lines. Higher spectral resolution observations are
needed to settle the question.}, nor the 125.4, 83.3, 66.4 and 58.7
$\mu$m water lines on-source.  As shown in our Fig. \ref{spectrum}, we
only detected the 179.5 $\mu$m 174.6 $\mu$m and 108.0 $\mu$m lines in
the outflow peak position NE-red, and the CO lines between J$_{up}$ =
14 and 17. We also do not confirm their detections of CO J$_{up}$
$\geq 18$ lines in the NE-red outflow peak position. Finally, in the
SW-blue position we only detected C$^+$ and OI 63 $\mu$m emission and
very marginally the H$_2$O 179 $\mu$m lines.
 

\begin{table*}
\label{fluxes}
\begin{center}
    \begin{tabular}{c c c c c c c}
	\hline
	\hline
	Specie&Transition&Wavelength&E$_{up}$&\multicolumn{3}{c}{Fluxes}\\
	&&($\mu$m)&(cm$^{-1})$&On source&NE-red&SW-blue\\
	\hline
	o-H$_{2}$O&2$_{21}$-2$_{21}$&180.49 & 134.9 &1.1 $\pm$0.5&$<$ 0.5&$<$ 0.5\\
	&2$_{12}$-1$_{01}$&179.53&79.5&2.7 $\pm$ 1.0&0.5 $\pm$0.4 & 0.6 $\pm$ 0.4\\
	&3$_{03}$-2$_{12}$&174.63&136.7&1.6 $\pm$ 0.7&0.5 $\pm$ 0.4 & $<$ 0.5\\
	&4$_{14}$-3$_{03}$&113.54&224.5&3.0 $\pm$ 1.0&$<$ 0.5&$<$ 0.5\\
	&2$_{21}$-1$_{10}$&108.07&134.9&2.0 $\pm$ 0.8&0.9 $\pm$ 0.5&$<$ 0.5\\
	&5$_{05}$-4$_{14}$&99.48&325.3&1.4 $\pm$ 0.6&$<$ 0.5&$<$ 0.5\\
	&6$_{16}$-5$_{05}$&82.03&447.3&1.2 $\pm$ 0.6&$<$ 0.5&$<$ 0.5\\
	&4$_{23}$-3$_{12}$&78.74&300.5&1.4 $\pm$ 0.6&$<$ 0.5&$<$ 0.5\\
	&3$_{21}$-2$_{12}$&75.38&212.1&1.7 $\pm$ 0.7&$<$ 0.5& $<$ 0.5\\
	\hline
	p-H$_{2}$O&3$_{22}$-3$_{13}$&156.19&206.3&0.5 $\pm$0.4&$<$ 0.5&$<$ 0.5\\
	&3$_{13}$-2$_{02}$&138.53&142.3&1.2 $\pm$ 0.6&$<$ 0.5& $<$ 0.5\\
	&2$_{20}$-1$_{11}$&100.98&136.2&1.0 $\pm$ 0.5&$<$ 0.5& $<$ 0.5\\
	&3$_{22}$-2$_{11}$&89.99&206.3&1.4 $\pm$ 0.6&$<$ 0.5& $<$ 0.5\\
	&3$_{31}$-2$_{20}$&67.09&285.1&0.6 $\pm$ 0.4&$<$ 0.5& $<$ 0.5 \\
	\hline
	CO&14-13&186.00&403.5&1.3 $\pm$ 0.6&1.2 $\pm$ 0.4&$<$ 0.5\\
	&15-14&173.63&461.1&1.5 $\pm$ 0.7&0.8 $\pm$ 0.4& $<$ 0.5\\
	&16-15&162.81&522.5&1.7 $\pm$ 0.7&0.7 $\pm$ 0.4&$<$ 0.5\\
	&17-16&153.26&587.7&1.7 $\pm$ 0.7&0.6 $\pm$ 0.5&$<$ 0.5\\
	&18-17&144.78&656.8&1.3 $\pm$ 0.6&$<$ 0.5&$<$ 0.5\\
	&19-18&137.20&729.7&1.3 $\pm$ 0.6&$<$ 0.5&$<$ 0.5\\
	&20-19&130.37&806.4&1.2 $\pm$ 0.6&$<$ 0.5&$<$ 0.5\\
	\hline
	OH&$^{2}\Pi_{3/2,5/2}$-$^{2}\Pi_{3/2,3/2}$&119.33&83.7&1.0 $\pm$ 0.5&$<$ 0.5&$<$ 0.5\\
	&$^{2}\Pi_{3/2,7/2}$-$^{2}\Pi_{3/2,5/2}$&84.51&201.9&1.1 $\pm$ 0.5&$<$ 0.5&$<$ 0.5\\
	&$^{2}\Pi_{1/2,1/2}$-$^{2}\Pi_{3/2,3/2}$&79.15&126.3&1.1 $\pm$ 0.5&$<$ 0.5&$<$ 0.5\\
	\hline
	[OI]&$^{3}$P$_{1}$-$^{3}$P$_{0}$&145.48&227.7&$<$ 0.5&$<$ 0.5&$<$ 0.5\\
	& $^{3}$P$_{2}$-$^{3}$P$_{2}$&63.17&158.7&2.2 $\pm$ 0.9&4.3 $\pm$ 1.5&2.2 $\pm$ 0.9\\
	\hline
	[CII]&$^{2}$P$_{3/2}$-$^{2}$P$_{1/2}$&157.74&63.7&2.2 $\pm$ 0.9&1.8 $\pm$ 0.7&1.1 $\pm$ 0.5\\
	\hline
    \end{tabular}
\end{center}
\caption{Measured line fluxes in units of 10$^{-12}$ ergs s$^{-1}$ cm$^{-2}$.
  Upper limits are given as 2 $\sigma$.}

\end{table*}

\section{Line modeling} 
\label{modeling} 
 
\subsection{Origin of the FIR line emission} 
\label{origin} 
 
The molecular emission (H$_2$O, CO and OH) observed toward IRAS4 can
have at least three different origins: the two outflows powered by
IRAS4A and IRAS4B, the PDR at the surface of the cloud, and the
collapsing envelopes around the two protostars.  The origin of the
molecular line emission can be disentangled when the spatial
distribution of the line emission and/or the line profiles are
available.  For example, lines arising in the envelope or in the
molecular cloud have narrow profiles whereas lines arising from
outflows show broadened profiles with extended wings.  Unfortunately,
in the case of the ISO-LWS observations, the relatively low spatial
and spectral resolution do not allow to observationally disentangle
the different components. However, the comparison between the central
and the NE-red and SW-blue positions allows a first guess of the
origin of the observed emission.  The [OI] and [CII] lines have
comparable line fluxes in the three observed positions.  For this
reason it is likely that the observed [OI] and [CII] emission is
associated with the ambient diffuse gas, either emitted in the PDR or
in the molecular cloud itself.  On the contrary, only the lowest lying
(J$_{up}$ $\leq$ 17) CO lines are detected on the NE-red position,
while no H$_2$O or CO emission is clearly detected on the SW-blue
position. On the other hand, the observations of the CO 3-2 line show
that the high velocity gas (the fastest outflow component) peaks at
the NE-red and SW-blue positions \citep{Blake95}. The lack of water
emission in these two outflow peak positions is not in favor of the
hypothesis that the on-source water emission originates in the
outflow. Although we cannot exclude a different origin and/or
contamination for example from the densest parts of the outflow
located in the ISO beam, in the following we explore the hypothesis of
the envelope thermal emission and interpret the observed water line
emission according to the CHT96 model.  The first goal of our modeling
is to verify that the thermal emission from the surrounding envelope
can reproduce the water line observations, a necessary condition even
though not sufficient to test this hypothesis.  A following section
will then address the possible origin of the CO and OH observed
emission.
 
\subsection{Model description} 
 
The CHT96 model computes in a self-consistent way the radiative
transfer, thermal balance, and chemistry of the main gas coolants
(i.e. O, CO and H$_2$O) across the envelope, in the \emph{inside-out}
framework \citep{Shu77}. Here we give a brief description of the main
aspects of the model.

The initial state of the envelope is assumed to be an isothermal
sphere in hydrostatic equilibrium, which density is given by:
 
\begin{equation} 
   \label{density_isothermal} 
   n_{\rm H_{2}}(r) = \frac{a^{2}}{2 \pi \mu m_{{\rm H}} G} r^{-2}
\end{equation} 
 
where $a$ is the sound speed, $m_{{\rm H}}$ is the hydrogen mass, $\mu$ is
the mean molecular mass, $r$ the distance from the center and $G$ the
gravitational constant.

At $t = 0$ the equilibrium is perturbed and the collapse starts from
inside out, propagating with the sound speed. The density in the inner
collapsing region is given by the free-fall solution:
 
\begin{equation} 
   \label{density_freefall} 
   n_{{\rm H_{2}}}(r) = \frac{1}{4 \pi \mu m_{{\rm H}}} \frac{\dot{M}}
   {(2GM_{*})^{1/2}} r^{-3/2}
\end{equation} 
 
where \Mstar is the star mass, \Mdot is the accretion rate,
related to the sound speed by:
 
\begin{equation} 
   \label{accretion_rate} 
   \dot{M} = 0.975 \frac{a ^{3}}{G}.
\end{equation} 
 
The free-fall velocity is given by:
 
\begin{equation} 
   \label{velocity} 
   v(r) = \left( \frac{2GM_{*}}{r} \right) ^{1/2}
\end{equation} 
 
The gravitational energy is released as material falls at the core
radius $R_{*}$, so that the luminosity of the protostar is:
 
\begin{equation} 
   \label{luminosity} 
   L_{*} = \frac{GM_{*}\dot{M}}{R_{*}}
\end{equation} 
 
In the following $L_{*}$ is the bolometric luminosity of IRAS4, and we
leave \Mstar and \Mdot as free parameters.
 
The radiative transfer in the envelope is solved in the escape
probability approximation in presence of warm dust, following the
\cite{Takahashi83} formalism.  The CHT96 model assumes that the
initial chemical composition is that of a molecular cloud, and then it
solves the time dependent equations for the chemical composition of 44
species, as the collapse evolves.  H$_2$O, CO and O are of particular
importance since they are the main coolants of the gas, and hence we
study the chemistry of these species in detail.  The CO molecule is
very stable, and its abundance results constant across the envelope.
H$_2$O is mainly formed by dissociative recombination of the
H$_{3}$O$^{+}$ in the cold outer envelope, while, at dust temperature
above 100 K, icy grain mantles evaporate, injecting large amounts of
water into the gas phase.  When the gas temperature exceeds $\sim$250
K, the H$_2$O formation is dominated by the endothermic reactions O +
H$_{2} \to $OH + H followed by H$_{2}$ + OH $\to$ H$_{2}$O + H, which
transform all the oxygen not locked in CO molecules into H$_2$O.
 
From the above equations and comments, the water line emission depends
on the mass of the central object, the accretion rate and the
abundance of H$_2$O in the outer envelope and in the warm region,
where its abundance is dominated by the mantle evaporation.  All these
quantities directly enter into the H$_2$O line emission, and
specifically into the determination of the H$_2$O column density.  In
fact, the accretion rate sets the density across the protostellar
region (Eqs.  \ref{density_isothermal} and \ref{density_freefall}).
The central mass of the protostar affects the velocity field, and
hence indirectly the line opacity (Eq. \ref{velocity}).  This
parameter also sets the density in the free-fall region
(Eq. \ref{density_freefall}) and therefore the gas column density in
this region. The water emission also depends indirectly on the O and
CO abundances, which enter in the thermal balance and hence in the gas
temperature determination.  Several recent studies \citep{Baluteau97,
Caux99, Vastel00, Lis01} have shown that almost the totality of the
oxygen in molecular clouds is in atomic form.  Accordingly, we assume
the oxygen abundance to be the standard interstellar value, i.e. $5
\times 10^{-4}$ with respect to H$_{2}$.  With regard to the CO
abundance, following \cite{Blake95} we adopt a CO abundance of
10$^{-5}$ with respect to H$_{2}$, lower than the standard abundance
as this molecule is believed to be depleted on IRAS4.  We will comment
later on the influence of these parameters on our results.  Finally,
the water abundance in the cold and in the warm parts of the envelope
are poorly known and are free parameters in our study.
 
To summarize, we applied the CHT96 model to IRAS4, and to reproduce
the observations we varied the four following parameters: the mass of
the central object \Mstar, the accretion rate \Mdot, the water
abundance in the outer cold envelope \Xout, and the water abundance in
the region of mantle evaporation \Xin.  The principal limitation to
the application of this model to IRAS4 is that the ISO beam includes
both IRAS 4A and IRAS 4B envelopes.  As a first approximation, we
assumed that the two envelopes contribute equally to the molecular
emission.  Finally we assumed that the two envelopes touch each other,
namely they have a radius of 3000 AU (i.e. $30''$ in diameter), in
agreement with millimeter continuum observations \citep{Motte01}.  In
our computations we adopted a distance of 220 pc in agreement with
JSD02 and a luminosity of 5.5 \solarlum for each
protostar, according to \citet{Sandell91} when assuming such a
distance.
 
\subsection{Water line modeling results} 
 
In order to constrain the mass, accretion rate and water abundance
across the envelope, we run several models varying the central mass
from 0.3 to 0.8 \solarmass, the accretion rate from 10$^{-5}$ to
10$^{-4}$ \solarmassyr, the water abundance in the outer parts of the
envelope \Xout between 10$^{-7}$ and 10$^{-6}$, and a water abundance
in the inner parts of the envelope \Xin between 10$^{-6}$ and $2
\times 10^{-5}$ respectively.  In the following we discuss the results
of this modeling.
 
\subsubsection{Accretion rate and water abundance in the outer parts 
   of the envelope} 
 
\begin{figure} 
    \resizebox{\hsize}{!}{ 
      \includegraphics[width=1.\columnwidth]{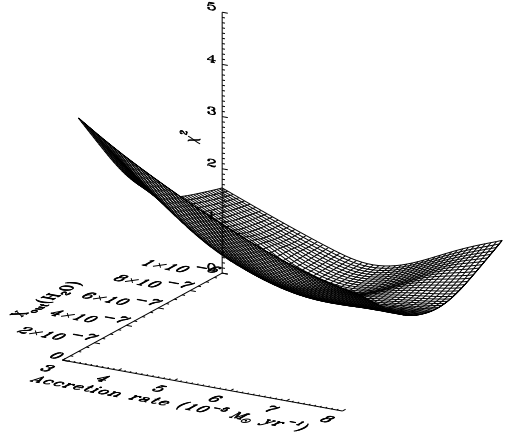}} 
    \caption{$\chi^2$ surface as function of the water abundance in
	the outer envelope \Xout and the mass accretion rate \Mdot.
	The $\chi^2$ has been obtained considering the lines with an
	upper level energy \Eup lower than 142 cm$^{-1}$ and for a
	central mass of 0.5 \solarmass.}
    \label{XoutMdot} 
\end{figure}

One of the difficulties in constraining the central mass, accretion
rate and water abundance in the envelope is that the water line
intensity a priori depends on all the parameters.  However, choosing
appropriate lines can help constraining one parameter at once.  Low
lying H$_2$O lines are expected to rapidly become optically thick in
the outer envelope, where they are easily excited.  Hence these lines
depend weakly on \Xin and \Mstar (which affect the line emission in
the collapsing inner region).  We therefore used the low-lying lines
to constrain the other two parameters, namely \Xout and \Mdot.  For
this we minimized the
$\chi^2=\frac{1}{N-1}\sum_{1}^{N}\frac{\left(Observations -
Model\right)^{2}}{\sigma^{2}}$ obtained considering only water lines
having a \Eup lower than 142 cm$^{-1}$, and where $\sigma$ is the
error associated with each line flux.  Fig.  \ref{XoutMdot} shows the
$\chi^{2}$ surface as a function of \Xout and \Mdot for $M_*$ = 0.5
\solarmass.  We obtained a similar plot for \Mstar = 0.3 \solarmass
and the result is basically the same.  As suspected, the chosen lines
constrain relatively efficiently the two parameters \Xout and \Mdot.
The minimum $\chi^2$ is obtained for a water abundance \Xout $\sim 5
\times 10^{-7}$ and an accretion rate \Mdot $\sim 5 \times 10^{-5}$
\solarmassyr.
 
\subsubsection{Central mass and water abundance in the inner parts of 
    the envelope} 
 
\begin{figure} 
    \resizebox{\hsize}{!}{ 
      \includegraphics[width=1.\columnwidth]{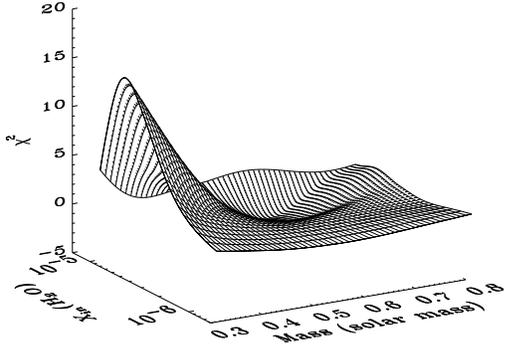}} 
    \caption{$\chi^2$ surface as function of the central mass \Mstar
      and water abundance in the innermost parts of the envelope \Xin.
      The $\chi^2$ has been obtained considering the lines with an
      upper level energy \Eup larger than 142 cm$^{-1}$ and assuming
      \Mdot = $5 \times 10^{-5}$ and \Xout = $5 \times 10^{-7}$. Note
      that we did not include the 82 and 99.5 $\mu$m lines, which
      seems underestimated by our model (see text). We did not
      include the 113 $\mu$m line either, because of the blending with
      the CO J$_{up}$ = 23 line, which makes the estimate of the flux
      rather uncertain.}
    \label{XevM} 
\end{figure}

We then constrained the central mass \Mstar and the abundance in the
innermost parts of the envelope \Xin using the high-lying lines.  In
fact to be excited, these lines require relatively high temperatures
and densities which are likely to be reached only in the innermost
parts of the envelope.  Their intensities depend hence on the water
abundance \Xin in these parts and on the central mass \Mstar.  The
$\chi^{2}$ surface as function of these two parameters is shown in
Fig. \ref{XevM}, obtained considering the lines having an upper level
energy \Eup larger than 142 cm$^{-1}$, and assuming \Mdot = $5 \times
10^{-5}$ and \Xout = $5 \times 10^{-7}$.  In this case the $\chi^2$
has a minimum around \Mstar = 0.5 \solarmass and \Xin = $5 \times
10^{-6}$. Specifically, if we adopt a constant H$_2$O abundance of
$5\times 10^{-7}$ ($X_{out}$) across the entire envelope, the model
predicts intensities a factor between two and five lower than those
observed (high lying lines, i.e. with \Eup $\geq 250$ K).  In other
words, the observed emission can only be accounted for if a jump in
the water abundance is introduced when the dust temperature exceeds
100 K, the sublimation temperature of icy mantles.  This jump has to
be larger than about a factor 10.
 
\subsubsection{The best fit model} 
 
Assuming two identical envelopes, the best fit model is obtained with a 
central mass of 0.5 \solarmass, accreting at $5 \times 10^{-5}$ 
\solarmassyr (Table 2). 

 \begin{table} 
   \begin{center} 
     \begin{tabular}{l c c} 
       \hline 
       \hline 
       Parameter & IRAS4 & IRAS16293-2422 \\ 
       \hline 
       Mass (\solarmass) & 0.5 & 0.8 \\ 
       Accretion rate (\solarmassyr) & $5 \times 10^{-5}$ & $3 \times
       10^{-5}$ \\ 
       Water abundance \Xout & $5 \times 10^{-7}$ & $3 \times 10^{-7}$\\ 
       Water abundance \Xin  & $5 \times 10^{-6}$ & $3 \times 10^{-6}$\\ 
       \hline 
       Radius (T$_{\rm dust}$ = 30 K) (AU) & 1500 & 4000 \\ 
       Radius (T$_{\rm dust}$ = 100 K) (AU) &   80 & 150 \\ 
       Age (yr)  & $1.0 \times 10^4$ & $2.7 \times 10^4$ \\ 
       \hline 
     \end{tabular} 
     \caption{Best fit parameters and derived quantities for IRAS4 and 
	comparison with the values obtained towards IRAS16293-2422 by 
	\citet{Ceccarelli00a}.} 
   \end{center}
  \label{bestfit}
\end{table} 

\begin{figure} 
    \resizebox{\hsize}{!}{
      \includegraphics[width=1.\columnwidth]{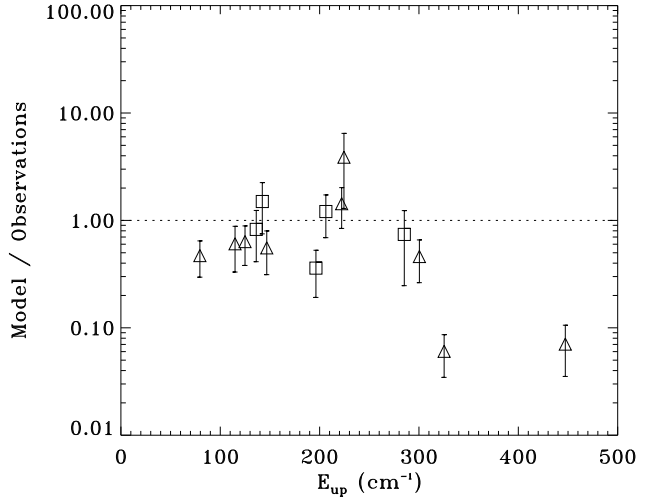}}
    \caption{Ratio between the line fluxes predicted by our best fit
      model and observed ones as function of the upper level energy of
      the transition \Eup.  Triangles represent ortho H$_2$O
      transitions, and squares represent para H$_2$O transitions.
      Note that the model assumes an ortho to para ratio equal to 3.
    }
    \label{ModelObs} 
\end{figure}

Assuming a constant accretion rate, this gives an age of 10000 years,
close to the dynamical age of the outflows.  The abundance of water in
the outer parts of the envelope is $5 \times 10^{-7}$ and it is
enhanced by a factor 10 in the innermost regions of the envelope,
where grain mantles evaporate.  Figure \ref{ModelObs} shows the ratio
between the observed and predicted line fluxes as function of the
upper level energy of the transition. The model reproduces reasonably
well the observed water emission, with the exception of the lines at
99.5 $\mu$m and 82.0 $\mu$m that seems to be underestimated (by a
factor 10) by our model.Since, on the contrary, lines in a comparable
range of \Eup are well reproduced by our model, we think that this
discrepancy is likely due to a rough baseline removal.  Specifically,
the estimate of both the 99.5 and 82.0 $\mu$m line fluxes may suffer
of an incorrect baseline removal, as the lines lie on the top of a
strong dust feature, which covers the 80-100 $\mu$m range (Ceccarelli,
Caux, Tielens et al. in preparation). Higher spectral resolution
observations are required to confirm this explanation. Finally some
unexplained discrepancy between the model and the observations may
exist at the higher values of \Eup.
 
\begin{figure} 
    \resizebox{\hsize}{!}{ 
      \includegraphics[width=1.\columnwidth]{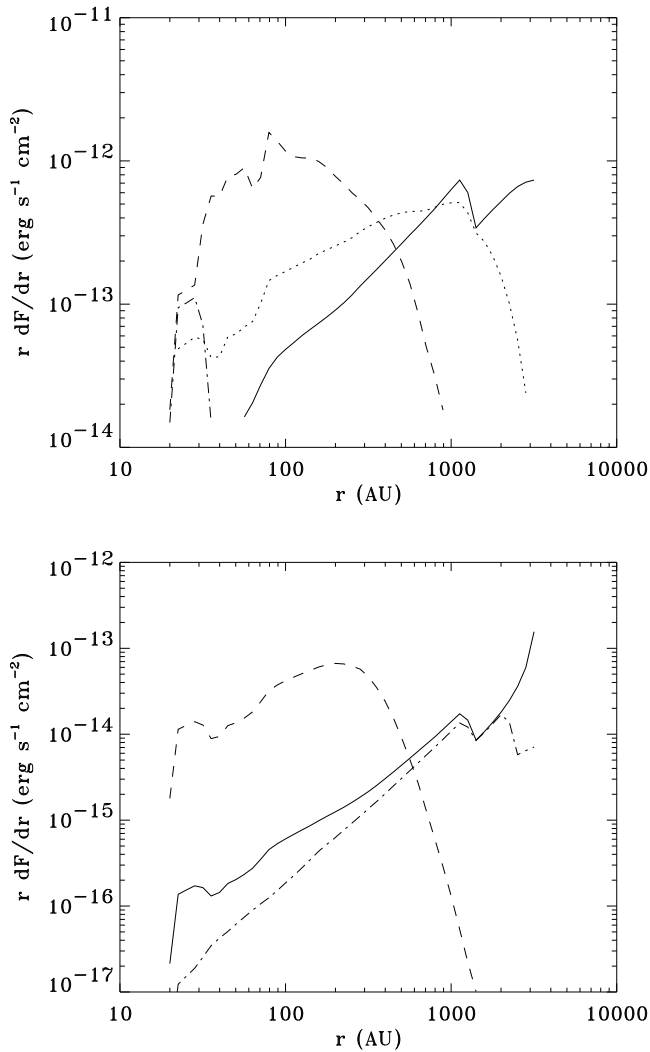}} 
    \caption{Predicted intensity of various lines as function 
      of the radius. The upper panel shows the water lines  at 179 
      $\mu$m (solid line), 108 $\mu$m (dotted line), 75 $\mu$m (dashed 
      line) and 82 $\mu$m (dash-dotted line). The lower panel shows the 
      CO J$_{up}$ = 3 (solid line), CO J$_{up}$ = 14 (dashed line) and 
      C$^{18}$O J$_{up}$ = 3 
      (dash-dotted line).} 
    \label{Profiles} 
\end{figure}

In the figure we also report different symbols for the ortho and para
water transitions respectively. In our model we assumed that this
ratio is equal to 3. The comparison between the observations and
predictions is consistent with this assumption. Plots of the predicted
intensity of various lines as function of the radius are reported in
Fig. \ref{Profiles}. Finally, the envelope model predicts an
intensity of 1.8 $\times$ 10$^{-13}$ erg s$^{-1}$ cm$^{-2}$ and a
linewidth of $\sim$ 1 Km s$^{-1}$ for the ground water line at 557
GHz, equivalent to an antenna temperature of 30 mK in the SWAS
beam ($\sim 4'$). SWAS detected a T$_a^* \sim$100 mK line, which linewidth 
is $\sim 18$ Km s$^{-1}$, and self-absorbed at the rest velocity (Bergin et
al. 2002). The observed line is undoubtedly dominated by the outflow
emission, and only a small fraction of its intensity can be attributed to
the envelope emission, in agreement with our model predictions.

\subsection{[OI], [CII], OH and CO emission} 
 
[OI] 63 $\mu$m and [CII] 157 $\mu$m emission is widespread, and
probably associated with the cloud.  A plausible explanation is that
the two lines are emitted in the PDR resulting from the UV and/or
X-ray illumination of this cloud from the several young stars that it
harbors.  The comparison of the observed fluxes with the model by
\citet{Kaufman99} suggests a PDR with a density of about 10$^{4}$
cm$^{-3}$ and a incident FUV of $\sim 5$ G$_{0}$.  This PDR would
account for the total observed flux of the [CII] 157 $\mu$m line and
OI.  The parameters we derive are in agreement with those quoted by
\citet{Molinari00}, who studied the region around SVS13.
 
The thermal emission from the envelope predicts no C$^+$ emission, of
course, as no source of ionization is considered in the CHT96 model.
The atomic oxygen, on the contrary, is present all along the envelope
and it is predicted to emit $1.8 \times 10^{-12}$ erg s$^{-1}$
cm$^{-2}$. This is similar to the observed [OI] 63 $\mu$m flux.  The
fact that we do not see any [OI] 63 $\mu$m enhancement towards the
source with respect to the surroundings can be explained if IRAS4 is
well embedded in the parental cloud. Being the ground transition, the
[OI] 63 $\mu$m line is relatively easily optically thick, and an
emission from an embedded source can be totally absorbed by the
foreground material \citep{Poglitsch96, Baluteau97, Caux99, Vastel00}.
 
Finally, our model predicts OH and CO J$_{up}$ $\geq$ 14 line fluxes
more than ten times lower than those observed. Note that the FIR CO
lines predicted by the CHT96 model are optically thick and not
sensitive to the adopted abundance, and therefore increasing the CO
abundance would not change the result. An extra heating mechanism is
evidently responsible for the excitation of the FIR CO lines observed
in the central position.  Shocks have been invoked in the literature
\citep[e.g.][]{Ceccarelli98, Nisini99, Giannini01}, but this
hypothesis has its own drawbacks and flaws (see Introduction).
Another possibility is that the FIR CO lines are emitted in a
superheated layer of gas at the surface of a flaring disk, as seen in
the protostar El 29 (Ceccarelli et al. 2002), and/or at the inner
interface of the envelope itself (Ceccarelli, Hollenbach, Tielens et
al. in preparation).  In the first case (disk surface), the gas is
``super-heated'' because the grain absorptivity in the visible exceeds
the grain emissivity in the infrared \citep[e.g.][]{Chiang97}.  In the
latter case (envelope inner interface) the extra heating is provided
by the X-ray photons from the central source. A full discussion
about the FIR CO emission origin is beyond the scope of this article
and we refer the interested reader to the above mentioned articles.
 
\section{Discussion} 
\label{discussion} 
 
\subsection{The structure of the envelope} 
 
\begin{figure} 
    \resizebox{\hsize}{!}{ 
      \includegraphics[width=0.9\columnwidth]{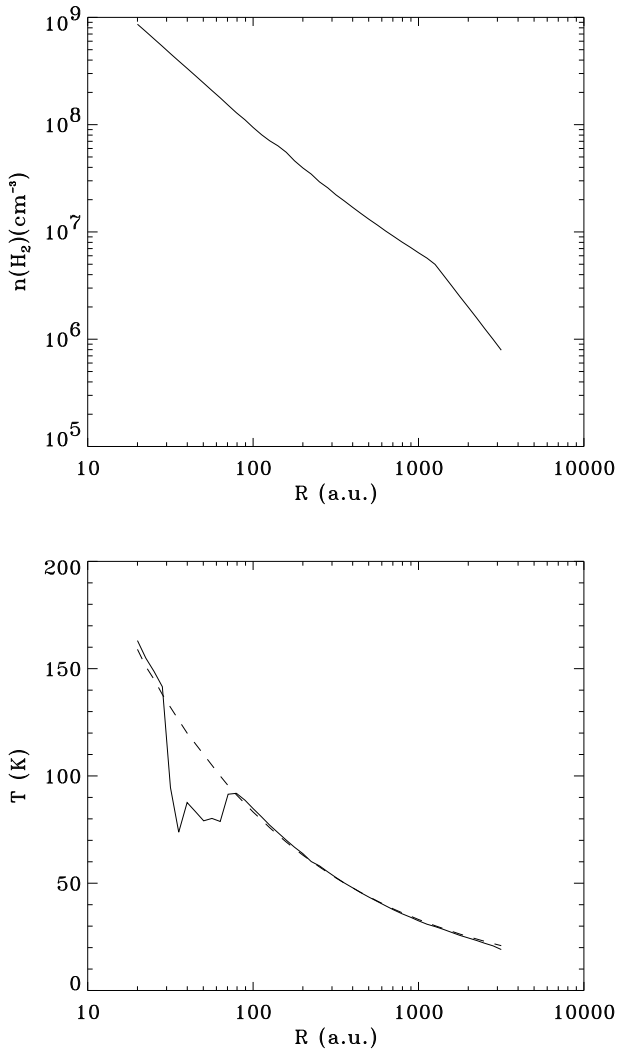}} 
    \caption{Density and temperature profiles of the envelope as 
      	computed by the best fit model. In the lower panel the dashed 
	line refers to the dust temperature, while the 
	solid line refers to the gas temperature. } 
    \label{dens-temp} 
\end{figure}

The first result of our modeling is its capacity to reproduce the
observed water emission.  In the following we show that the derived
parameter values are in agreement with previous estimates obtained
through different studies.  Before discussing in detail this
comparison we describe here the derived density and temperature
profiles, as well as the gas heating and cooling mechanisms.  The
reconstructed density and temperature profiles are shown in
Fig. \ref{dens-temp}.  In the outer region the density follows a
$r^{-2}$ law (static region), while in the inner region the density is
proportional to $r^{-3/2}$ (free-fall region), with a transition
region starting at $\sim 1500$ AU, that covers a substantial range of
radii and is flatter than $r^{-3/2}$. The absolute density is
relatively large, $4 \times 10^6$ cm$^{-3}$ at 1500 AU.
 
The gas temperature closely follows the dust temperature in the outer
and intermediate regions of the envelope. At $r \sim 80$ AU, dust and
gas decouple. This is caused by the evaporation of the icy mantles
when the dust temperature reaches 100 K, which injects large amounts
of water in the gas phase (about a factor ten more) increasing the
cooling efficiency of the gas.  The gas temperature drops by about 20
K and remains lower than the dust temperature until at $r \sim 30$ AU
the dust FIR pumping of the water molecules counterbalances the water
line cooling and couples again the gas and dust temperatures.  The
heating and cooling of the gas in the envelope follow the general
properties discussed in CHT96 \citep[see also][]{Ceccarelli00a}.  The
heating is dominated by the NIR water pumping, gas-grains collisions
and the compression of the gas in the innermost regions, while
gas-dust collisions are the main heating factor in the outer parts of
the envelope and when the dust and gas decouple at $r \sim 80$ AU.
The gas cooling is dominated by the line emission of water, oxygen and
CO. In the outer parts of the envelope the cooling is dominated by the
CO line emission, while in the intermediate region the cooling is
dominated by the oxygen and water lines, as the CO lines become
rapidly thick.  At $r \leq 80$ AU, the cooling by the water lines
takes over, due to the icy mantle evaporation, and dominates the
cooling by orders of magnitude with respect to O and CO. Note that
before the ices evaporation (i.e. at r $\ge$80 AU), the heating is
dominated by the gas compression and the cooling by the H$_2$O line
emission.  The increase of the gaseous water abundance by a factor ten
causes an increase of about the same amount in the gas cooling rate,
whereas the compression heating rate just increases by 20\%.  The gas
heating becomes hence dominated by the gas-grains collisions, which
tend to couple dust and gas. In the specific case of IRAS4, the water
cooling rate $\Lambda_{{\rm H_{2}O}}$ and gas-grains collisions heating
rates $\Gamma_{{\rm dg}}$ can be approximated by the following expressions,
at the radius just before the evaporation (for the details see CHT96):

\begin{eqnarray}
  \Lambda_{\rm{H_{2}O}} &=& 3.4 \times 10^{-15} ~\frac{x({\rm H_2O})}{5
  \times 10^{-7}} ~\left( \frac{\rm T_{gas}}{90 {\rm K}} \right)^{3/2}
  {\rm erg~s^{-1} cm^{-3}}\\ 
  \Gamma_{\rm{dg}} &=& 3.6 \times 10^{-16} ~({\rm T_{dust} - T_{gas}})
  ~\left( \frac{\rm T_{gas}}{90 {\rm K}} \right)^{1/2} {\rm erg~s^{-1}
  cm^{-3}}
\end{eqnarray}

The increase of a factor ten in the water cooling rate after the ice
evaporation is only in partly counter-balanced by the increased
heating rate due to the gas-grains collisions, and the difference
between the two temperatures is $\sim 20$ K.
 
The CO and atomic oxygen abundances are constant across the envelope,
within the studied range, i.e. 30 to 3000 AU.  We will discuss in the
next paragraph the effect of varying the CO abundance across the
envelope to take into account the CO depletion when the dust
temperature is below the CO-rich ice evaporation temperature.  As
widely discussed previously, the water abundance undergoes a jump of
about a factor ten at $r \leq 80$ AU, when H$_2$O-rich ices evaporate
(dust temperature larger than 100 K).
 
One interesting prediction of this study is the existence of a hot
core like region in the innermost parts of the envelope, where the
dust temperature reaches the sublimation temperature of the grain
mantles.
 
\subsection{CO depletion} 
 
Here we want to address in some detail the issue of the CO depletion
claimed in IRAS4 by comparing low-lying, millimeter CO transitions and
dust continuum emission \citep{Blake95}. Blake et al. used the CO J =
3-2 transitions of the $^{12}$C, $^{18}$O and $^{17}$O isotopes and
found that the emission is accounted for a CO abundance of $\sim
2.5\times 10^{-5}$, i.e. a factor five lower than the ``canonical''
value.  Their explanation is that CO molecules freeze out on the grain
mantles and the gas phase CO results therefore depleted across the
envelope.  This result has been recently confirmed by JSD02, who quote
CO/H$_2 \sim 10^{-5}$ in the outer envelope of IRAS4. The question
arises whether the CO is depleted across the entire envelope or not,
as also remarked by JSD02.  We checked with our model if the
millimeter observations would be sensitive to higher abundances of CO
in regions where the dust temperature exceeds the CO-rich ice
evaporation temperature, i.e. $\sim 25-30$ K.  We found that if the
C$^{18}$O J = 3-2 line is optically thin, the bulk of the J = 3-2 line
is emitted at $\sim 3000$ AU. This is because of the excitation of the
CO J = 3-2 line itself.  In the optically thin and thermalized line
approximation it is possible to derive the following approximate
analytical expression (strictly speaking, valid only for the CO
millimeter lines):
 
\begin{equation} 
I_J \propto \int \frac{{\rm exp}[-E_J/k{\rm T_{gas}}]}{\rm T_{gas}} dr 
\end{equation} 
 
When J = 3 the integrand has a peak around ${\rm T_{gas}} \sim 30$ K,
does not vary by more than 20\% with temperatures between 20 K and 40
K, and decreases for temperatures larger than $\sim 40$ K.  As a
result, the line intensity in the outer envelope is linearly
proportional to the radius of the emitting gas, and therefore
increases going outward, i.e. the emission is dominated by the outer
envelope where CO is depleted.
 
We run a case in which there is no CO depletion across the envelope,
(i.e. the CO abundance is constant and equal to $10^{-4}$), a case
where CO is depleted on the entire envelope (i.e. the CO abundance is
constant and equal to $2.5 \times 10^{-5}$), and a third case where CO
is depleted only in the outer envelope (i.e. the CO abundance is $1
\times 10^{-5}$ when ${\rm T_{dust}} \leq 30$ K and $10^{-4}$ in the
rest of the envelope).  The last two cases give the same intensity
(within 15\%) on both C$^{18}$O J = 3-2 and J = 2-1 lines, while the
first one gives an intensity larger by about a factor 3.  Therefore,
the last two cases, CO depletion across the entire envelope and
depletion in the outer envelope only, are not distinguishable by the
C$^{18}$O J = 3-2 and J = 2-1 observations.  In this sense, we think
that the $J_{up} \leq 3$ observations cannot probe the innermost
regions and that it is possible that the measured CO depletion is only
relative to the external regions of the envelope.

This situation would be similar to what has been claimed to occur
in IRAS16293-2422 \citep{Ceccarelli01}, based on the indirect evidence
provided by the D$_2$CO emission.  The D$_2$CO molecule is considered
a grain mantle product, as gas phase reactions seem unable to form an
appreciable amount of this molecule.  Ceccarelli et al. showed that in
IRAS16293-2422 the D$_2$CO emission originates in the region of the
envelope where ${\rm T_{dust}} \geq 30$ K.  The proposed
interpretation is that D$_2$CO is trapped in CO-rich ices that
evaporate when the dust temperature exceeds 30 K.  Hence, in
IRAS16293-2422 there is an outer region of the envelope where CO is
frozen onto the grain mantles (${\rm T_{dust}} \leq 30$ K), and a
regions with ${\rm T_{dust}} \geq 30$ K where CO is released into the
gas phase and has the standard $\sim 10^{-4}$ abundance.  A similar
scenario has been also suggested by JSD02 for other
Class 0 sources that show CO depletion.
 
\subsection{Comparison with previous studies of IRAS4} 
 
In this paragraph we compare our results with previous studies dealing
in a way or in another with some of the issues addressed in the
present study.  We start with the recent study by JSD02, who derived
the density profile of the IRAS4 envelopes by modeling the continuum
emission (spectral energy distribution and 450/850 $\mu$m maps
simultaneously) and assuming a single power law index.  They found a
power law index equal to 1.8 and 1.3 for 4A and 4B respectively,
consistent with the Shu inside-out solution adopted in our model.  At
1000 AU they estimate a density equal to 6 (4A) and 2 (4B) $\times
10^{6}$ cm$^{-3}$, which are quite comparable with our estimate: $6
\times 10^{6}$ cm$^{-3}$.  We emphasize that the two methods, their
and our, are totally independent, and use different data, continuum
and line observations respectively.  The fact that both predict
approximately the same density structure in a certain way validate
both methods, or at least increases the probability that both models
describe reasonably well IRAS4.
 
Apart from the density and temperature profiles, our model also constrains
the water abundance profile.  It is reasonable to ask whether our
predicted water abundance in the innermost and outer regions of the
envelope are realistic and if they have any support from different
observations.  The situation here is somewhat complicated by the fact
that there aren't many other independent ways to measure the water
abundance.  From a theoretical point of view the abundance in the
outer envelope, $5 \times 10^{-7}$, can be very well compared with chemistry
model predictions \citep[e.g. ][]{Lee96}.  In this respect, the value
that we derive is certainly not extraordinary and rather plausible.
From an observational point of view Bergin et al. (2002)
succeeded to detect the 557 GHz water line in the NGC1333 molecular
cloud.  They estimate the water abundance in the region to be $\sim
10^{-7}$, with unfortunately a relatively large error ($\sim 10$) due
to the many uncertainties in the excitation of the line.
\citet{Moneti01} derived a water abundance of $3 \times 10^{-7}$ in
the clouds in the line of sight of the galactic center. These authors
claim that this is very likely the abundance of standard molecular
clouds. In summary, the water abundance that we find for the cold
region of the IRAS4 envelope is consistent with other studies.
Regarding the abundance in the inner region, $5 \times 10^{-6}$, the
value that we obtain seem to be lower than what expected if all the
water ice is injected in the gas phase and a large fraction of the
oxygen is locked in this ice.  A typical water ice abundance is
estimated around $10^{-4}$ \citep{Tielens91}. However, SWAS
observations of IRAS4 and other low mass protostars suggest that the
water abundance in their outflows is around $10^{-6}$ (Neufeld et
al. 2001; Bergin et al. 2002), i.e. similar to the value that we find.
Those estimates are very rough and could easily be off by a factor ten
\citep{Neufeld00}, as they are based on one transition only, but
nevertheless have the advantage that the observed emission is
certainly dominated by the outflow (the spectral resolution of these
observations is $\sim$ 1.2 km s$^{-1}$) so there are no doubts on its
origin.  Since the water abundance in the outflow would be probably
dominated by the grain mantles released in the gas phase, these
observations probably measure the water content in the mantles, very
much as our observations measure (indirectly) the water content
mantles in the inner hot like region.  The two measurements seem to be
consistent in giving a rather low value.  Whether this validates both
measures is less certain than the density profile case: it certainly
does not discredit the two measures.  Finally, even the comparison of
our estimate of the accretion rate and central mass are in good
agreement  with the previous estimates, based on a different method (line
profile and molecule H$_2$CO), by \citet{DiFrancesco01}.  We derived
\Mdot = $5 \times 10^{-5}$ \solarmassyr against the $11-4 \times
10^{-5}$ \solarmassyr quoted by \citet{DiFrancesco01}, and \Mstar =
0.5 \solarmass against the 0.23 - 0.71 \solarmass.
 
To conclude, these studies show that the values we derive of the four
parameters of our model are plausible and nothing of particularly
surprising, with the possible exception of the water abundance in the
innermost regions.  In other words, if we had to choose {\it a priori}
those values we would have chosen exactly what we found.  The
conclusion is that it is very probable that at least most of the
observed water emission in IRAS4 originates in the envelopes.  If any,
just a small fraction should therefore be associated with the outflow.
Our final comment is therefore that care should be taken when
interpreting the observed water emission towards low mass, Class 0
protostars as due to shocks (e.g. Ceccarelli et al. 1998, Nisini et al
1999, GNL01), as we showed in two out two cases that the massive
envelopes surrounding these sources dominate the water emission, just
because of the large total column density.  As a matter of fact, Class
I sources, which are characterized by less massive envelopes, do not
show up strong water emission \citep{Ceccarelli00a}.

\subsection{Comparison with IRAS16293-2422} 

The mass and accretion rate we derived for IRAS4A and B are of the
same order of magnitude of those found in IRAS16293-2422
\citep{Ceccarelli00a}.  IRAS16293-2422 seems more massive (0.8
\solarmass) than IRAS 4A (0.5 \solarmass), and accreting at a slightly
lower accretion rate (3 against 5 $\times 10^{-5}$ \solarmassyr).
Assuming a constant accretion rate, those values give an age of 10000
years and 27000 for IRAS4 and IRAS16293-2422 respectively.  Hence
IRAS16293-2422 seems more evolved than IRAS4.  Moreover, IRAS4
possesses an hot core like region about two times smaller than
IRAS16293-2422 (80 AU against 150 AU).  Ground-based H$_2$CO and
CH$_{3}$OH observations (Blake et al. 1995; Maret et al. in
preparation) confirm that IRAS4 is in fact colder, and therefore less
bright in these molecular transitions than IRAS16293-2422, and that
indeed the IRAS4 hot core like region is very small.  This fact
coupled with the larger distance of IRAS4 from the Sun may explain the
apparent difference in the molecular emission of these two sources,
which is much richer in IRAS16293-2422.  This conclusion is also in
agreement with the relatively higher millimeter continuum observed in
IRAS4, which implies a larger amount of cold dust surrounding this
source than IRAS16293-2422.  In addition, the region where the dust
temperature is higher than 30 K is smaller in IRAS4 ($\sim 1500$ AU)
than in IRAS16293-2422 ($\sim 4000$ AU), i.e. the CO depleted part of
the envelope is relatively larger in IRAS4 than in IRAS16293-2422.
This may explain why the CO depletion has been observed towards IRAS4
and not in IRAS16293-2422 \citep{vanDishoeck95,Ceccarelli00b}.
 
Finally, despite this difference in the age, the water abundance in
the envelope is remarkably similar in the two sources, both in the
outer part of the envelope and in the inner ones, where ice mantles
are predicted to evaporate.  This is an important piece of
information, suggesting that the ice mantle formation in the two
sources underwent a similar process, despite the macroscopic
difference between the two molecular clouds which the two sources
belong to. In the case of IRAS16293-2422, the cloud seems very
quiescent, shielded from strong UV and/or X-ray radiation
\citep[e.g.][]{Castets01} and with even evidence of large CO depletion
\citep{Caux99}.  In the other case, IRAS4, the cloud presents cavities
excavated by the several young stars of the region
\citep[e.g.][]{Lefloch98}, and it is probably permeated by the X-rays
emitted by them. A forthcoming study will allow to measure the H$_2$CO
and CH$_{3}$OH abundances in the inner hot core like region of IRAS4
(Maret et al. in preparation) and make comparisons with that found in
IRAS16283-2422 \citep{Ceccarelli00b}. This study will hence help to
understand in more detail how apparently different conditions in the
parental clouds affect the grain mantle composition.
 
\section{Conclusions} 
 
We presented a spectral survey of the protobinary system IRAS4 in the
NGC1333 cloud, using ISO-LWS in grating mode.  We targeted the source
as well as two adjacent positions, NE-red and SW-blue, that encompass
the red and blue lobes of the outflow emanating from IRAS4,
respectively.  The three spectra are dominated by the [OI] 63 $\mu$m
and CII [157] $\mu$m lines, that likely originate in the PDR
associated with the parental cloud. On the contrary, water emission is
only detected towards the on-source position, whereas no significant
water emission is detected towards the NE-red and SW-blue positions.
This suggests that the bulk of the water emission is due to the
thermal emission of the protostellar envelopes around the two
protostars.  Using an accurate model of the chemistry, thermal balance
and radiative transfer in protostellar envelopes (CHT96), we modeled
the water line emission due to two identical envelopes surrounding
IRAS4A and B respectively. We found that the observations are
consistent with the CHT96 model, which implicitly assumes the
``inside-out'' theory \citep{Shu77}.  The best fit of the model allows
us to estimate the four main model parameters: the accretion rate, $5
\times 10^{-5}$ \solarmassyr, the central mass, 0.5 \solarmass, the
water abundance in the outer envelope, $5 \times 10^{-7}$, and in the
inner envelope, $5 \times 10^{-6}$ (this last parameter is the least
constrained with about a factor 3 of uncertainty).  This gives an age
of 10000 years, assuming that the accretion rate remains constant
during the collapse.  Based on this model, we derived the density and
temperature profiles of the gas in the envelopes.  We also reviewed
the suggestion by \citet{Blake95} that CO is depleted by about a
factor ten in the envelope of IRAS4.  We could not confirm or rule out
this hypothesis but caution that the transitions used by this study,
C$^{18}$O 3-2 and 2-1, can hardly probe the inner regions, where the
CO abundance may be ``canonical''.
 
A comparison with several previous studies of the same source (Blake
et al (1995), Neufeld et al 2000, DiFrancesco et al. 2001, JSD02)
shows that the derived parameters are reasonable and consistent with
the available literature, hence re-enforcing the thesis that the
observed water emission is indeed due to the thermal emission from the
envelopes.  A by-product of the present study is the prediction of the
existence of a hot core like region in the inner parts of the
envelope, where grain mantles evaporate, releasing large amounts of
water (about a factor ten) in the gas phase. Such a hot core has
already been proposed to exist around IRAS16293-2422, where a similar
study as been carried out \citep{Ceccarelli00a, Ceccarelli00b}.
Comparison between the two protostars, show that IRAS4 is younger and
surrounded by a more massive envelope.  This explains the larger
continuum emission and the larger depletion factors observed in IRAS4.
Finally, this study emphasis the necessity of ground based
observations, where higher spatial and spectral resolutions are
achievable. H$_2$CO and CH$_3$OH are of particular interest as they
are among the most abundant components of grain mantles, and are
therefore expected to evaporate in the inner parts of the
envelope. Appropriate transitions can hence be used to constrain the
physical and chemical conditions in the innermost part of protostellar
envelopes \citep[see][]{Ceccarelli00b}.
 
\begin{acknowledgements} 
We wish to thank Edwin A. Bergin for frank and constructive
discussions on the SWAS data.  We thank Edwin A. Bergin and Jes K. J\o
rgensen for providing us with their papers prior to publication. The
referee Neal Evans is thanked for his useful comments. Most of the
computations presented in this paper were performed at the Service
Commun de Calcul Intensif de l'Observatoire de Grenoble (SCCI).
\end{acknowledgements} 
 
\bibliography{2736} 
\bibliographystyle{apj} 
\nocite{*}

\end{document}